**Intermittent *in-situ* high-resolution X-ray microscopy of 400-nm porous glass under uniaxial compression: study of pore changes and crack formation**

Sebastian Schäfer[1], François Willot,[2] Mansoureh Norouzi Rad,[3] Stephen T. Kelly,[3] Dirk Enke,[4] Juliana Martins de Souza e Silva[1,5,*]

[1] Institute of Physics, Martin-Luther-Universität Halle-Wittenberg, Halle, Germany

[2] Mines ParisTech, PSL - Research University, Centre for Mathematical Morphology, Fontainebleau, France

[3] Carl Zeiss X-ray Microscopy, Pleasanton, CA, USA

[4] Institute of Chemical Technology, Universität Leipzig, Leipzig, Germany

[5] Fraunhofer Institute for Microstructure of Materials and Systems IMWS, Halle (Saale), Germany

**Abstract**

The properties of porous glasses and their field of application strongly depend on the characteristics of the void space. Understanding the relationship between their porous structure and failure behaviour can contribute to the development of porous glasses with long-term reliability optimized for specific applications. In the present work, we used X-ray computed tomography with nanometric resolution (nano-CT) to image a controlled pore glass (CPG) with 400 nm-sized pores whilst undergoing uniaxial compression *in-situ* to emulate a stress process. Our results show that *in-situ* nano-CT provides an ideal platform for identifying the mechanisms of damage within glass with pores of 400 nm, as it allowed the tracking of the pores and struts change of shape during compression until specimen failure. We have also applied computational tools to quantify the microstructural changes within the CPG sample by mapping the displacements and strain fields, and to numerically simulate the behaviour of the CPG using a Fast Fourier Transform/phase-field method. Both experimental and numerical data show local shear deformation, organized along bands, consistent with the appearance and propagation of ± 45 degrees cracks.

**Introduction**

Porous silica monoliths can be synthesized by phase separation processes, either through the thermally-induced phase separation in alkali borosilicate glasses or the polymer-induced phase separation in sol-gel processes [1-3]. They are used in optics, heterogeneous catalysis, sensors, chromatography, and as hosts for nanoparticles and bio-compounds mainly due to their high specific surface area and high thermal and chemical resistances [1, 3]. The fine-tuning of the synthetic parameters allows tailoring the silica pore network, resulting in monomodal or hierarchical pore structures with pore size distribution ranging from nano- to micrometer scales [1, 3].

For use in any of those contexts, the mechanical stability of porous silica is an essential requirement to guarantee resilience, load-bearing capacity, and fatigue resistance [4]. However, regardless of their potential, porous silica monoliths' poor mechanical stability hinders their widespread use at an industrial scale. Questions remain open concerning the predictability of their mechanical properties, particularly the onset of local damage and its propagation throughout the volume until collapse. *In-situ* three-dimensional (3D) X-ray imaging has already been used in materials science for following structural evolution over time, whether during manufacturing, throughout service, or in understanding the events leading to failure [5]. It enables the detection of large [6] to sub-micrometric pores [7], and was used to image mechanical tests in laboratories and synchrotron facilities during *ex-situ* and *in-situ* indentation and compression [8-10]. When combined with Digital Volume Correlation (DVC) analysis, *in-situ* X-ray CT imaging data can produce three-dimensional images of the undeformed and deformed states. A three-dimensional grid defined in the reference (undeformed) volume allows calculating the displacement for each point of this grid [11]. Experimentally estimated local variations in the microstructure, displacement fields, and strain fields enables quantitative study the internal deformation and damage process of materials[12], and can help establishing potential routes to explain failure mechanisms [8]. Indeed, the DVC method enables to solve the problems of microstructural deformation and crack initiation and expansion in materials [12].

The observation of spatial and temporal damage occurrences is critical for the understanding of a materials behavior [13]. Here we investigated a porous glass sample with interconnected 400 nm-sized pores under *in-situ* uniaxial compression with high-resolution X-ray imaging. We evaluated the pores shape changes, microdamage formation, and evolution, in addition to volumetric residual strain distribution due to uniaxial compression using DVC. Besides that, we used a phase-field formulation to numerically predict crack nucleation and propagation in this material. To the best of the authors knowledge, direct quantitative comparisons in 3D between numerical models of crack initiation/propagation and experimental characterization in heterogeneous media are not available so far for macroporous silica with sub-micrometric pores.

**Experimental Section**

*Sample preparation and laser machining*

Porous glass monoliths with 400 nm pores (pore size estimated by mercury intrusion porosimetry, in Fig. S1) were prepared by thermal treatment of an initial glass with the composition (mol.%) 62.5 $SiO_2$, 30.5 $B_2O_3$ and 7 $Na_2O$ at 700 °C for 24 h using a Nabertherm electric furnace to induce a spinodal phase separation in the material. The formed sodium-rich borate phase was subsequently removed by leaching in 1 M HCl for 2 h at 90 °C. The CPG monoliths were further leached in 0.5 M NaOH for 2 h at room temperature to remove secondary colloidal silica from the pores [3, 4]. After each leaching step, the monolith was washed with deionized water and dried. A disk was cut from one CPG monolith and immobilized onto a support for laser machining in a microPREP™ (3D-Micromac AG, Chemnitz, Germany) and a conical-shaped pillar of approximately 300 μm height and a tip of 40.05 ± 1.32 μm in diameter was laser-machined on the top of a cubic base, which was then glued on the tip of a pin using super glue.

*Intermittent in-situ compression test in X-ray microscope*

Imaging experiments were performed in a Carl Zeiss Xradia 810 Ultra operating with a chromium X-ray source (5.4 keV) using absorption contrast. Four full tomography datasets were obtained, each with an exposure time of 1 s per projection, and a total of 901 projections over a reduced angular range (± 70°). The first tomography was obtained from the reference state of the sample loaded to a ZEISS Xradia Ultra Load Stage comprising a piezo-mechanical actuator, a force gauge, and a pair of flat anvils between which the sample is mounted. Absolute force and absolute displacement were recorded during the entire experiment, enabling the estimation of the engineering stress, $\varepsilon$ ($\varepsilon = (l - lo)/lo$, with $l$ being the specimen current height, and $lo$ the initial height). After the first tomography obtained without any compression, the second to fifth full tomographies were obtained after compression increments of 3.8, 10.0, 15.7 and 19.4%, respectively, corresponding to $\varepsilon = -0.05$, $\varepsilon = -0.15$, $\varepsilon = -0.23$, and $\varepsilon = -0.28$, respectively.

*Image processing and Digital Volume Correlation*

Image reconstruction was performed by filtered back-projection algorithm using the software XMReconstructor integrated into the Xradia 810 Ultra. Final 16-bit images had a voxel size of 128 nm. The commercial software Avizo (Thermo Fisher Scientific, version 3D 2022.2) was used for image correction, segmentation, digital volume correlation analysis, and 3D renderings presented here. The datasets contrast was matched using the mean grey value of the uncompressed dataset as reference. A subvolume of 250×250×300 voxels was extracted from each dataset to remove borders but keeping the largest volume available. The images were processed to separate the solid phase from the porous phase by first applying a median filter (1 iteration), followed by interactive threshold segmentation. To define the value for the threshold cut-off for all datasets, the dataset of the uncompressed sample was segmented to provide the same porosity value estimated by mercury intrusion, equal to 58% (Fig. S1), as in [14]. After segmentation, the sample volume fraction for each xy-slice in each dataset was estimated, which corresponds to the ratio of the number of pixels corresponding to the specimen (glass) to the total number of pixels (glass and air) in the slice. Avizo Digital Volume Correlation (DVC) algorithm (global approach) was run with a cell size of 15 μm (defined after the uncertainty measurement from 6 μm with an increment of 2 for 12 steps [8]) and a convergence criterion of 0.001.

Phase-field predictions of failure

An FFT-based phase-field method was employed to simulate numerically uniaxial compression on the CPG400 sample. The same subvolume as used for DVC containing 270×280×310 voxels was considered, representing a domain of volume 35×36×40 μm$^3$. The local properties were chosen by inverse analysis. It is noted that in the linear elastic regime, the compression test was consistent with a macroscopic Young's modulus of 40 MPa. We first fixed the Poisson ratio in the solid phase to $\nu_s$ =0.2. The FFT computations carried out in the linear elastic regime show that, due to the high porosity, the effective (i.e. macroscopic) Young modulus $E_m$ of the structure is about 25% that of the solid phase $E_s$, so that $E_m \approx 0.25 E_s$. This relationship holds to a good accuracy independently of the Poisson ratio, in the range 0.1-0.3 (not shown). Therefore, we choose $E_s$=160 MPa and $\nu_s$ =0.2 in the solid phase. To predict brittle fracture numerically, i.e. crack nucleation and propagation, we used the phase-field formulation of Bourdin et al. [15] and Miehe et al. [16] of the variational principle introduced by Francfort and Marigo [17]. More specifically, a Fourier-based method was implemented [18-20], where the problem is decoupled into an elasticity and a phase-field systems of partial differential equations (PDE). The phase-field is a smooth scalar function in between 0 and 1 that serves to describe local damage, which is updated after each loading step. A key numerical parameter is the length-scale, which monitors the spatial fluctuations of the phase-field, and the width of the crack. In the present work, it was set to two voxels. Another important property is the "unilateral law" that governs the crack in the damaged zone, and here the law proposed by N'Guyen et al [21] was used. In addition, a sequential algorithm with fixed loading increments of 0.01% so that 2000 loading steps are necessary to reach a compression of 20% was used. The critical energy release rate was chosen so that a macroscopic crack occurs at a strain of about 20%. For comparison, a second volume of size 350×300×200 voxels taken from another specimen of the same CPG400 sample imaged in another instrument was considered. Convergence was monitored by two convergence criteria for the strain and phase-field problems, which are treated separately at each loading step. To mimic simple compression along the y-direction, we use mixed loading conditions so that $\varepsilon_{zz}<0$, and $\sigma_{xx}=\sigma_{yy}=0$.

**Results**

We performed full 3D scans of the CPG400 sample in a total of five times: step 0 is the initial state before any compression and 1 to 4 are consecutive vertical (z-axis) compression steps applied while controlling the sample displacement in the z-axis (Fig. 1). Step 4 is the last one and lead to failure of the sample. The force-displacement curve (Fig. 1 a) shows sudden drops in the force, occurring in increments of 50-60 mN. Smaller drops repeat roughly every 10 mN between steps 1 and 3. The total force required for specimen failure is about a factor of ten smaller than for the non-porous counterpart (Fig. S2). With the dataset's segmentation, voxels were attributed to the sample, and the remaining ones were attributed to air and the volumetric representation of the specimen imaged after each compression (Fig. 1 b-f) show its changes of the shape, with a lateral dislocation after the second and third compressions (Fig. 1 d, and e) and a large crack after the fourth compression (Fig. 1 f).

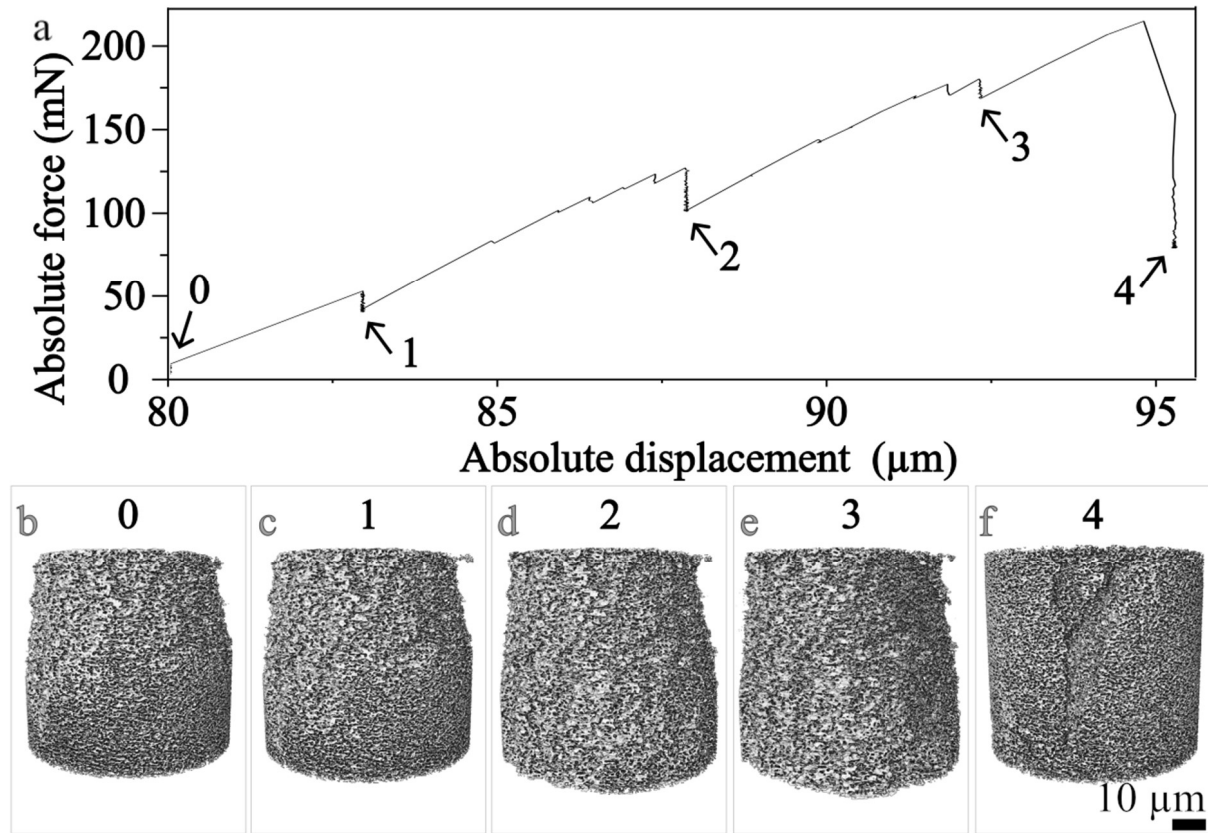

**Figure 1.** *In-situ* nano-CT characterization of CPG400 sample. a) Load-displacement curve for incremental compression. Pauses for CT scans were taken at the stress drops marked in the graph, b-f) CPG400 volumetric reconstructions for uncompressed sample and after incremental compressions (0: uncompressed sample; 1, 2, 3, and 4: after successive compressions).

A comparison of virtual slices in the same position before and after the compressions show changes occurring in the specimen in a smaller size scale (Fig. 2 a-j). By tracing one specific region on the top of the specimen (Fig. 2, blue arrows within red and blue squares), we observe the vertical movement of specific areas and their shape change. In another region, we see the complete closure of cavities on the top of the specimen (white arrows in Fig. 2 a-c within red squares, and Fig. 2 f-h) and a slight densification in the top layer (regions within white dotted lines in Fig. 2) The densification on the top of the specimen as an effect of the compression intensifies after the second and third compressions. Also, a crack appears at ±45° after the third compression (Fig. 2 d, and i, red arrow), which after the fourth compression seems to open wide in a crack in the border of the specimen (Fig. 2 e) but close in the middle of it, probably due to chipping and collapse of a chunk of specimen (Fig. 2 j). After the fourth compression, cracks open at ±45° in different position (Fig. 2 j, red arrows) and, besides the densification already observed on the top of the specimen basically after each compression state, a densification band appears close to the top also at ±45º (Fig. 2 f, white dotted lines).

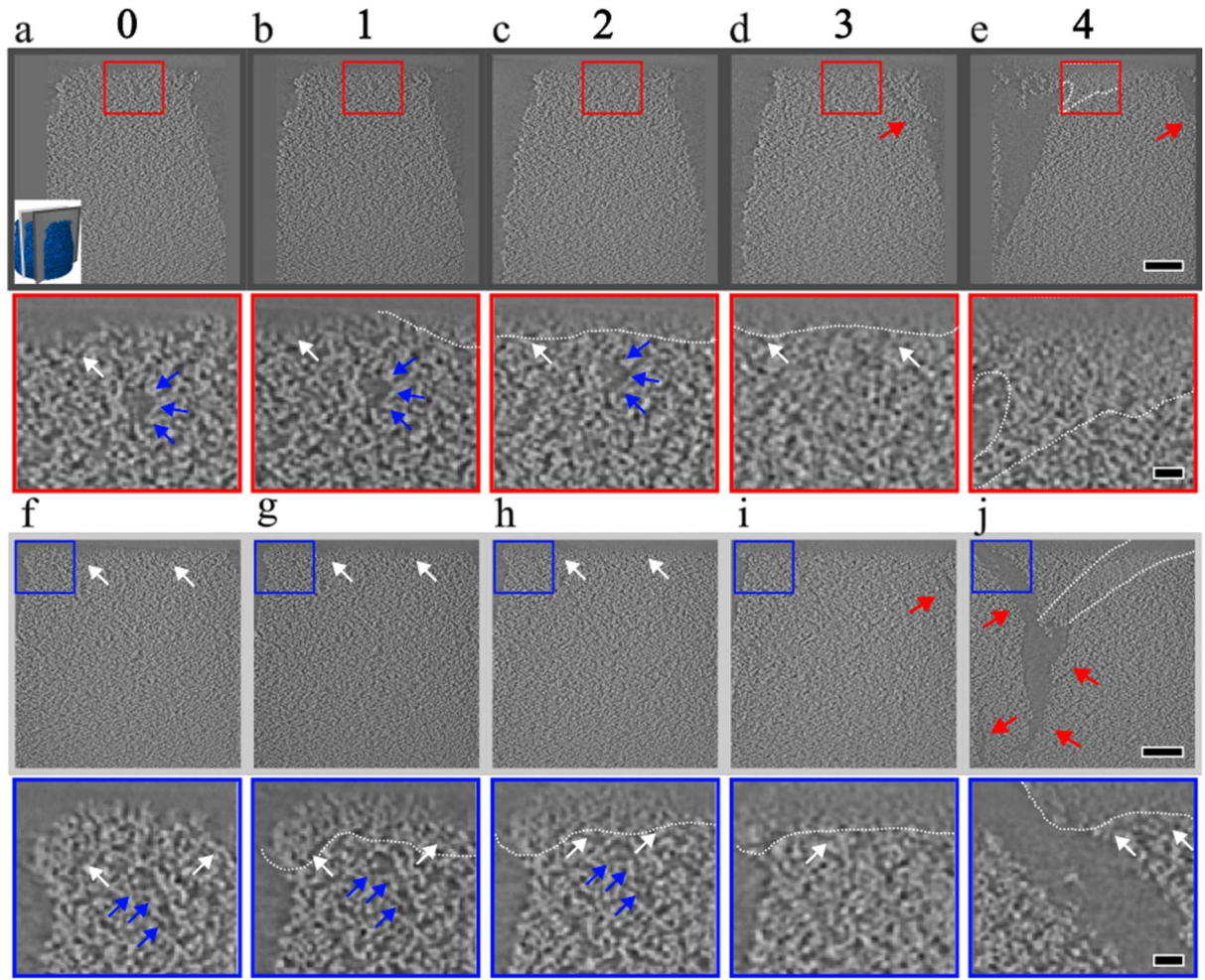

**Figure 2.** CPG400 vertical virtual slices (a-e) close to the border and (f-j) in the middle of the specimen, for uncompressed sample (0) and after four incremental compressions (1-4). Volumetric image in the detail in a) shows the position of the virtual slices in the specimen. Areas within the red and blue squares are shown enlarged immediately below. White arrows indicate areas with sample densification, red arrows indicate areas with cracks and sample detachment, blue arrow indicate same point of the specimen moving due to compression. Scale bars: 20 µm, and 2 µm (red and blue squares). White dotted lines define densification area.

A smaller volume of the specimen shows the pores in three dimensions (Fig. 3 a-e) and the shape of the large crack formed after the fourth compression (Fig. 3 e). Enlarged areas from an inner part of those volumes show in more detail pores and struts changing in shape after incremental compression. We followed the morphology in specific regions of the specimen to illustrate those changes microscopically (Fig. 3 f-j). All features followed move up in a diagonal fashion after each compression state. When following the compression effect in one strut (Fig. 3, within the yellow line), successive compressions resulted in the strut thickening (Fig. 3 f – h), branching after the third compression (Fig. 3 i), and collapsing after the fourth compression (Fig. 3 j). Another strut (Fig. 3, within the blue line) buckles after the first compression (Fig. 3 g) and breaks after the second (Fig. 3 h). A pair of pores (Fig. 3, within red line) slightly change in shape with the first compression (Fig. 3 g). Then, the top pore reduces in size after the second compression (Fig. 3 h) and the pores seem to merge after the third compression (Fig. 3 i).

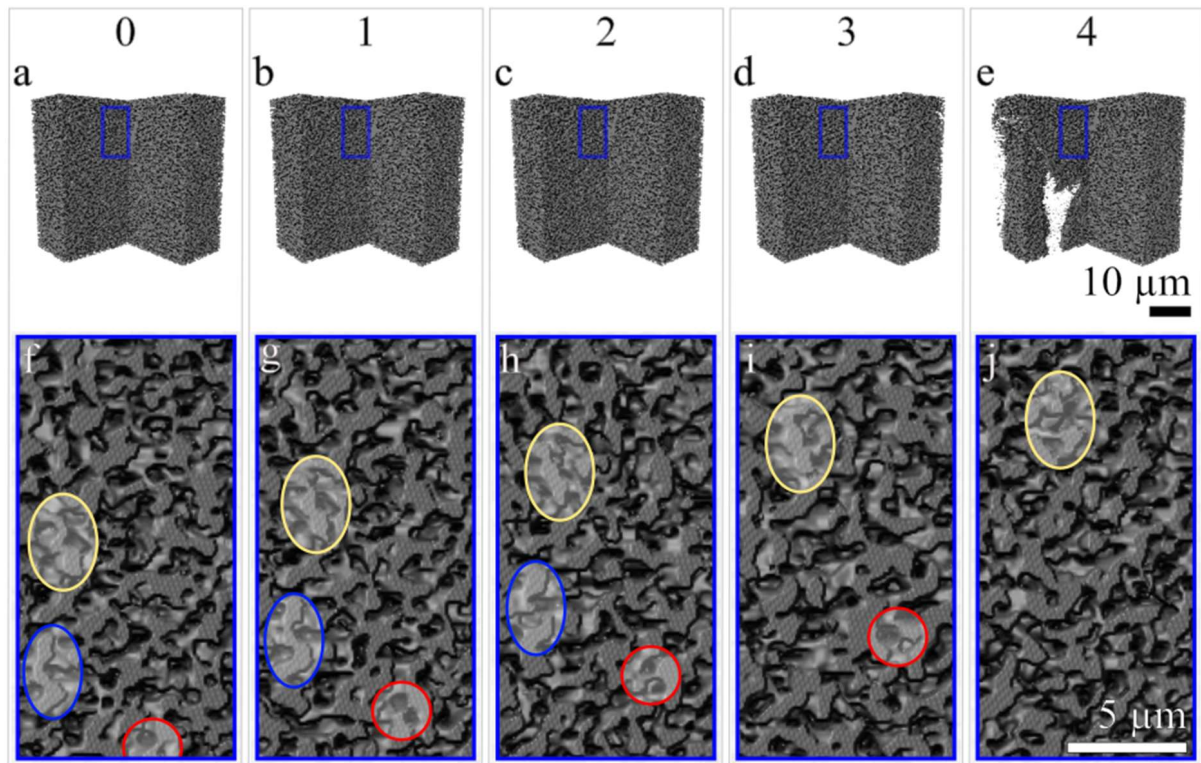

**Figure 3.** Volumetric rendering of a small volume of interest of the CPG400 specimen. (a) uncompressed and (b-e) after successive first to fourth compressions. Area highlighted in blue is shown immediately below (f-j), with pores and struts highlighted within blue, red, and yellow areas.

The global mechanical behaviour of material is linked to the evolution of local strain heterogeneities [11]. To visualize the local strains changes that take place due to the applied deformation, we used the displacement vectors superimposed on the local strain field obtained by DVC for the CPG400 (Fig. 4). The CPG400 exhibits a heterogeneous strain field. Higher strain is observed diagonally (Fig. 4 a-c) and reaches its highest values in the middle of the specimen with the third compression (Fig. 4 g), when the first crack is observed (Fig. 2). Lowest strain values are observed after the fourth compression (Fig. 4 d), which leads to the collapse of the specimen The vectors show the direction of the motion of the specimen during the successive compressions, which occurs in an angle (Fig. 4 e-h).

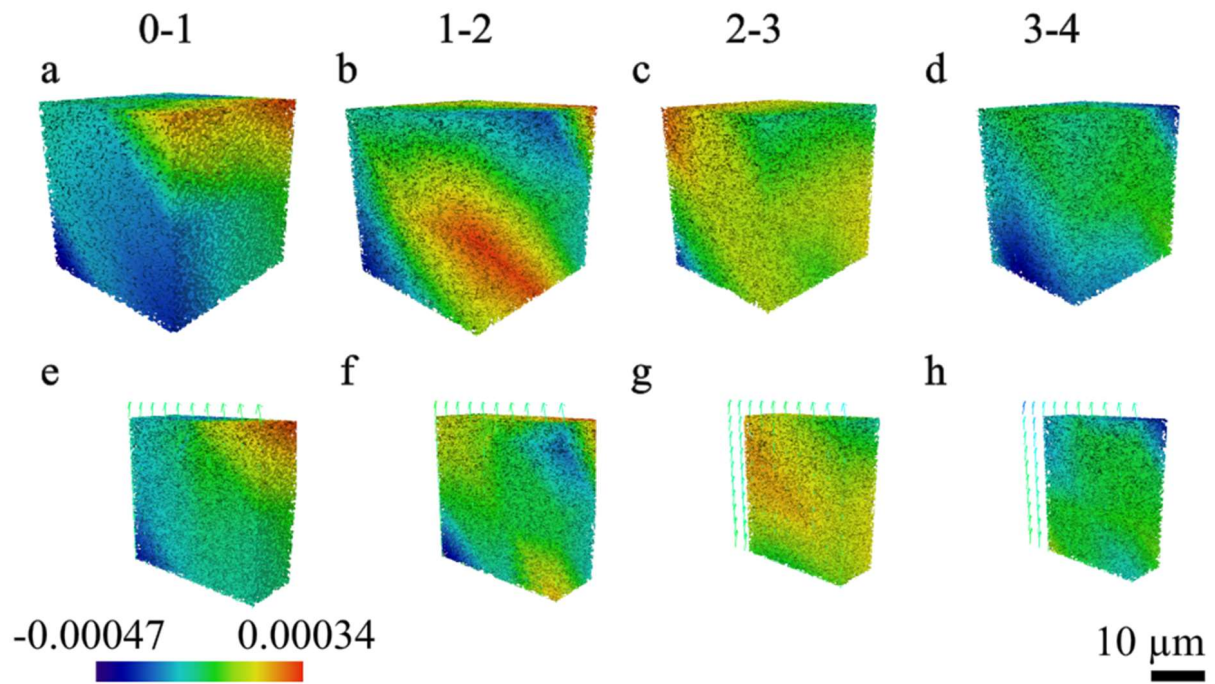

**Figure 4.** Strain field (Ezz) of the CPG400 shown in Fig. 3 (250×250×300). Data computed between a) uncompressed and first compression, b) first and second, c) second and third, d) third and fourth compressions, all superimposed to the 3D image of the previous step. Displacement vectors shown for same slice in all images.

To numerically predict brittle fracture (crack nucleation and propagation), we used the phase-field formulation. The stress-strain macroscopic response for the two specimens considered in our analysis have quite similar profile (Fig. 5 a). For one specific slice (Fig. 5 b), FFT maps equivalent strain and phase field (Fig. 5 c and d, respectively) obtained for the final loading step show that the cracks are oriented at angles of ±45° with respect to the loading direction.

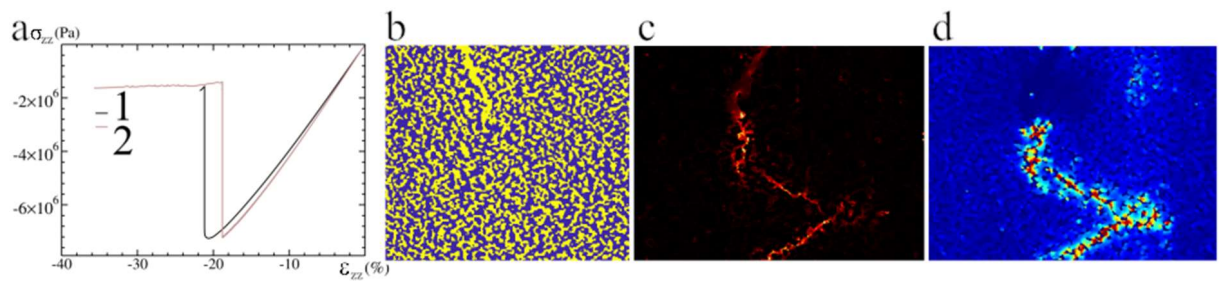

**Figure 5.** Phase-field predictions of failure. a) Simulated strain-stress curve for two specimens of the same sample, b) binarized slice xy=184, c) FFT predictions of the von Mises equivalent

strain (highest values in yellow, lowest in black) along a 2D-cut of slice xy=184 and d) corresponding phase-field φ (φ=1: red, damaged material; φ=0: blue, sane material).

A comparison of the 3D representation of the crack numerically predicted (Fig. 6 a) after nearly complete failure and the one obtained experimentally (Fig. 6 b and c, respectively for all specimen and a smaller volume selected within it) show that, though the crack is not located in the exact same region as the actual final experimental crack visible, is has a similar orientation.

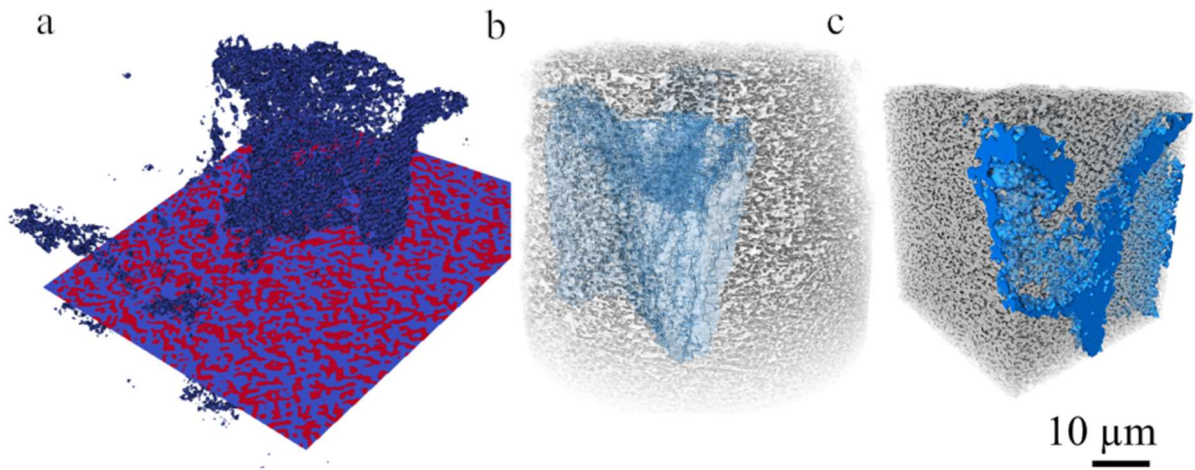

**Figure 6.** Three-dimensional view of the crack formed after a) phase-field simulation, and experimental compression crack in blue superimposed on the b) volumetric image and c) on a smaller work volume of the uncompressed specimen.

**Discussion**

The design of longer-lasting porous materials requires material failure examination and damage mechanisms identification. In general, damage originates at the nanoscale, accumulates, and grows until microscopic and macroscopic defects are formed [13]. Load-displacement curves describe the structural changes averaged statistically over the total specimen volume measured and, thus, assume stress–strain relationships to be homogeneous throughout the sample, neglecting the complex modification of internal pore morphologies due to local changes [22]. Besides the load-displacement curves, *in-situ* nano-CT allows for observing damage mechanisms at high resolution. Here, the nano-CT images of a 50μm-diameter CPG400 specimen under incremental loading allowed for visualizing individual pores and their spatial arrangement along successive compression states. It revealed that a force of 150 mN caused the appearance of the first crack in the specimen analysed, and further compression up to 200 mN caused specimen collapse. Microscopically, CPG400 pores change shape, opening and closing in different points, and struts curve, thicken and break after incremental compression, as seen in the two- and three-dimensional nano-CT images obtained (Figs. 1 – 3). Virtual slices show that the first crack observed formed under compression grows into a larger crack after further compression (Fig. 2 d, and e), and densification bands are observed in the vicinity of larger cracks.

We used DVC to estimate 3D displacements, which were then translated into strain fields. The strain maps and displacement vectors show that the changes occur heterogeneously within the specimen and that the CPG400 exhibits a heterogeneous strain field, which may be due to its microstructure character [11]. As observed by others [11], with increasing the load, local strain heterogeneities decreased (Fig. 4 b-d), indicating that the areas strongly deformed at the beginning of loading are less deformed afterward due to a re-localization of the imposed strain. The combined Fourier/phase-field method seems suitable for the numerical simulation of the behavior of the CPG400 subjected to uniaxial compression. The results obtained indicate the presence of local shear deformation along bands, possibly oriented at ±45° degrees, as observed experimentally. This mode of deformation, with varying angles, has been observed in other materials like metallic glasses [23].

Under uniaxial vertical strain, changes in the pore volume and the porosity are functions of the changes in vertical overburden stress [24]. As the load increases, the existing pores may coalesce, increasing in volume, but also collapse, causing densification of the specimen. Although nano-CT could provide accurate measurements of individual struts of the CPG400 under compressive load and allowed for observing changes in individual pores (Fig. 2), the resolution of the images limits significantly the ability to detect in detail the reduction of the pores' size and, thus, changes in the specimen's porosity due to compression. While the virtual slices allow for visualizing the sample densification on the top of the specimen after each compression state and the formation of densification regions formed as bands (Fig. 2), this information is lost with the segmentation protocol used (Fig. S3), making the quantification of the volume fraction inaccurate. As in [14], the threshold used for segmentation was chosen so that the volume fraction of the 3D image had a porosity as close as possible to the known porosity of the specimen measured by the standard method. This choice seems adequate for phase-field simulations and DVC calculations performed, as it enabled the detection of the pores, their interconnection, and local variations in the CPG400 microstructure. However, due to the resolution limit of the images obtained, the used segmentation protocol does not detect the smaller pores formed due to compression of the sample, and the information of the specimens' densification either on the top of the specimen or even within it is mostly lost.

Material failure examination and damage mechanisms identification are critical in the design of longer-lasting porous materials. Repeated experiments would be necessary to obtain statistically relevant data to confirm the forces necessary to cause this sample's damage. Moreover, it is well known for brittle materials that specimen size influences the mechanical properties measured, and the underlying mechanics at different length scales which govern brittle materials' behaviour are yet to be fully understood [25, 26]. This paper does not touch on the size subject and is only concerned with local structural changes in the CPG400 porous glass.

**Conclusion**

Here we investigated the local deformation of 400 nm-size controlled pore glass under mechanical stress events using *in-situ* uniaxial compression with high-resolution X-ray computed tomography, Digital Volume Correlation, and FFT numerical modelling. Our results show that the high-resolution 3D images effectively captured the structural deformation occurring as an effect of the vertical compression of the specimen. They illustrate pore densification, crack initiation, propagation, and failure of the sub-micrometric pore silica studied. Analysis of different areas of the specimen revealed the internal deformation occurring during compressive stress, including the dynamic changes of specific pores and struts in three-dimensions. Through the images, it is possible to detect densification areas, although higher-resolution imaging may be necessary to accurately quantify the densification at the top of the specimen or in bands within it. Results suggest that the crack propagation mechanisms were controlled by pre-existing cracks formed with a similar angle.

Without any a priori knowledge about the location of the initiation of the cracks, the combined Fourier/phase-field method seems suitable for the numerical simulation of the CPG400 and shows that the cracks formed due to compression are oriented at angles of ±45° to the loading direction. The results obtained indicate the presence of local shear deformation along bands, possibly oriented at ±45° degrees, as also observed experimentally. Repeated experiments would be necessary to obtain statistically relevant data regarding the relation of stress-strain in this sample, though it remains unclear if the small-scale experimental and numerical results accurately capture the damage propagation mechanisms in large specimens.